# COMPARATIVE ANALYSIS OF 3D-CNN MODELS, GARCH-ANN, AND VAR MODELS FOR DETERMINING EQUITY PRICES

**Sydney Anuyah**[1] **Mary I. Akinyemi**[2] **and Chika O. Yinka-Banjo**

[1]Department of Computer Science, University of Lagos, Nigeria
[2]Department of Statistics, University of Lagos, Nigeria

(sydneyanuyah@gmail.com, Makinyemi@unilag..edu.ng, cyinkabanjo@unilag.edu.ng,)

**Abstract**
Financial models have increasingly become popular in recent times, and the focus of researchers has been to find the perfect model which fits all circumstances; however, this has not been thoroughly achieved, and as a result, many financial models have been created. Artificial Intelligence modelling has increasingly become more popular in the financial space as an answer to the weakness of the advanced mathematical models studied in Economics. This paper introduces three commonly used models and tests them on the S&P500 to give a strong projection as to the future values of the prices. It then introduces various error metrics like the root mean square error (RMSE) to ascertain the viability of the models. The results show that a longer-term forecast indeed has more arduous consequences as there is a veer between the actual and the forecasted readings. The models can produce a strong growth estimate for businesses and policymakers to plan more appropriately in the short term. Overall, the GARCH-LSTM model produced the least RMSE which also shows that complex models tend to perform better than simpler models

# 1.0 INTRODUCTION

Time series is the attribution of a variable that changes over time to the different seasons in which these changes occur. (Li et al., 2002) Examples of time series data include the amount of daily rainfall, wind speed, foreign exchange rate, equity prices, etc. Public equity is one financial instrument that has gained traction over the years with many companies in Nigeria like MTNN, BUACEMENT, Nestle, etc., offering their stock prices to the public. The financial market is an effective one and as such, the laws of demand and supply balance out effectively with a price change. Judging from the data available to the public provided by the National Association of Securities Dealers Automated Quotations (NASDAQ) every time, prices of stocks that are listed on the stock exchange, during active trading hours are seen to fluctuate regularly, usually with less than a minute interval. This behaviour of the market shows the market efficiency and shows how the law of demand and supply affects asset prices. When there is a high demand, prices go up to balance out and when there is a high supply (people selling their stocks), the prices go down to commensurate with that effect.

Citing an example, an individual joined the stock market recently and searched for the highest return traded stocks and saw an instrument that had a 15% increase in the last five days. The person purchased the stocks due to past trends and the stock skyrocketed a whopping 14% in three days. The person immediately sold the equity and made large profits. The excitement made the person try the same method with another instrument but this time, the person lost 22% of their invested funds, as these instruments were highly volatile. Now, what if such a person had the power of foretelling, maybe seeing into the future, they will probably not purchase the second instrument despite the rising bubble. This is the aim of this project, taking different models and comparing them to find out the individual forecasting strengths, by using past trends, to make educated guesses of what is to come to reduce the effect of random gambling.

## 1.1. AIM OF THE STUDY

Over the years, financial engineers and economists have used the pattern of equity to provide educational guesses on the future of stock prices. Models like Auto-Regressive (AR) models and Moving Average (MA) models have been used to create equations to predict future prices. Furthering the development of these models an integration between the two models gave rise to the Auto-Regressive Integrated Moving Average model (ARIMA also called ARMA). Due to the failure in accuracy especially in long-term forecasting, the Auto-Regressive Conditional Heteroscedasticity (ARCH) model was created. Due to the technicalities of the ARCH model, the Generalized Auto-Regressive Conditional Heteroscedasticity (GARCH) was introduced, which was meant to generalize the ARCH model. Further studies have shown that Recurrent Neural Networks also work well in predicting prices.

The failure of these models was largely due to uncontrolled circumstances. For instance, during the 2008 housing bubble, the stock market crashed and the real estate, building, and infrastructure companies lost the greatest value. Before this sudden crash, financial models predicted a consistent boom that went against the true value of the stock prices. However, the introduction of another variable made a great difference. This model was tagged vector autoregressive model which is used to provide another relatable variable to determine the future prices of stock. It engages products that are highly correlated, usually two rival products e.g. Coke and Pepsi, Nike and Reebok, etc. For example, a law passed on carbonated drinks will affect both Coke and Pepsi

## 1.2. FORECASTING MODELS

GARCH model is popular in the financial sphere and it is used to experiment with various macroeconomic variables. (Kristjanpoller & Minutolo, 2015, Kurniasari et. al., 2023) This model has been used in forecasting prices of financial instruments with different levels of volatility. Shareholders, Economic researchers, Analysts, Traders, Portfolio managers, etc., have always tried ceaselessly to invest in the highest gaining stocks based on different analyses. Many models have been used to improve the detection of stock prices in the past like the autoregressive (AR) model, the moving average (MA) model and the combination of the AR and the MA model to produce the Autoregressive integrated moving average model (ARIMA), the ARCH model which is an improvement from the AR model and the GARCH model which evolved from the ARCH model.

Gupta (Gupta et al., 2020) used the autoregressive model to predict international equity prices. This model worked well and was attributed to the fact that a reasonable value of p was used based on correlograms. The success of the model was majorly attributed to understanding the data in question first before predictive analysis began.

The moving average (MA) model is just like every other mean average taken over a sample of time. The moving average of a model is calculated using the total number of prices within a time allocation and divided by the number of total periods.

Wei (Wei et al., 2014) proposed a moving average model hybrid to predict the stock returns of TAIEX. The MA model was combined with a system called the "adaptive neuro-fuzzy inference system" or "adaptive network-based fuzzy inference system" (ANFIS) to provide more accurate results. It is common to see a lot of combinations of models e.g., ARIMA, in the financial space to ensure that we get the best outcome for our returns.

Che (Che & Wang, 2010) also employed the ARIMA model coupled with the support vector regression technique which is a machine-learning algorithm to also predict the prices of stock. The ARIMA is a combination of the autoregressive and the moving average. Now, coupled with the introduction of the support vector machine, it further proves that various combinations are possible. These combinations compensate for what each of the individual models missed in forecasting.

## 1.3. ARCH/GARCH MODELS

The GARCH Model in its raw form has been used to often try to ascertain the prices of instruments like equity. Tim Bollerslev (Bollerslev, 1987) described asset prices and interest rates in time-series data to show a level of non-correlation with time as asset prices do not always remain tranquil and experience different volatility levels at different times similar to the Brownian motion established by Bachelier. To the untrained eye, the asset prices are unstable and left to guesswork, but Chou (Chou, 1988) explained how the GARCH model can provide us with a thorough estimate of the value of an asset price in the market. According to Arnerić, (Arnerić et al., 2014) the GARCH model is the most widely used method for predicting volatility. This is because they account for the asymmetrical effects of high-frequency data sets, volatility clustering, time- varying volatility, and excess kurtosis.

Forecasting asset prices is not limited to equity alone, as financial instruments in the commodity market like gold, copper, electricity, etc., also make use of the GARCH model (Khosravi et al., 2013; Yan-ping, 2007). Even though this model accounts for the varying volatility of asset prices, usually, asset prices are influenced beyond the value of what the model accounts for. The GARCH model according to (McMillan & Speight, 2004) does not do well outside the sample data fed to it, because of the failure to measure the "true volatility". The "true volatility" here refers to the swing in asset prices in the real world. This inaccuracy leads to a forecast error defeating the goal of forecasting which is to eliminate the random error as much as possible. When the random error becomes too large, the forecast model is poor and it is sometimes as good as guessing or sometimes even worse than guessing as illustrated by Theil's U statistic. (Chand et al., 2012).

To improve the accuracy of forecasts, researchers have employed neural network applications in their work. In (Liu & So, 2020), Wing used the combination of the artificial neural network (ANN) with the GARCH (1, 1) algorithm to produce a model which surpasses the typical GARCH (1, 1) model. Yan (Hu et al., 2020), integrated the long short term memory (LSTM) neural network algorithm to predict copper prices. Yan further experimented with merging the neural network algorithms to get the best accuracy. This improvement in forecasting can be attributed to the neural system. (Dos Santos Coelho & Santos, 2011). The advantage of the neural network is that it can detect complex nonlinear attributions between the price and its variance, to properly predict future price values. (Kumari 2023) also used a special kind of ANN called RNN to predict banana prices

Much work has been done in predicting price models, however, in this article, we will be delving into the scedastic function which is an important variable in the GARCH model. This variance is calculated given certain assumptions that will be factored in by the neural network algorithm. The use of recurrent neural networks in forecasting over feedforward is explored in this paper, this is because the output usually learns better based on the feedback mechanism introduced in the network. (Arnerić et al., 2014). The error of the GARCH model from the actual is forecasted by the LSTM and used to provide a basis for the random noise which is expected from the model.

## 1.4. VECTOR AUTOREGRESSIVE MODELS

Vector Auto-Regressive Model is an offshoot of the Auto-Regressive model. The model was made popular around the 1980s. The Vector autoregressive model uses the popular illustration of apples and oranges. Consider a departmental store in which there are just two products sold, i.e. apples and oranges. A smart economist can use the correlation of the sales of the apples to derive the expected value of the sale of oranges will be in the new month. Since it is an autoregressive model, the previous values predict the future values. Now the VAR model combines both variables of apples and oranges and studies their correlation to hopefully arrive at a better and more accurate model than the simple AR model because it increases the number of variables in consideration.

In (Giudici & Abu-Hashish, 2019), a VAR model was used to measure the behavior of various cryptocurrencies and how they all interact with each other in the real world. Using the VAR model, the author attempted to forecast the down and upswings of the extreme volatile product. With this research, the author was able to provide more insights into the predictability of the volatility of cryptocurrency. Thus, allowing the analyst to understand the possibility of making cryptocurrency a viable investment opportunity.

In (Munkhdalai et al., 2020), there was an integration to improve the reliability of the VAR model just like the GARCH model. In this paper, the VAR model was integrated with the Gated Recurrent Unit (GRU) which is a new modification of the Neural Network Systems and bears semblance to LSTM.

### 1.5. 3D-CNN MODELS

Neural Networks have increasingly become more popular in recent times. This is because the concept of neural networks is established based on the patterns that have been observed in the brain. A 3D CNN Model also called a 3- dimensional convolutional neural network has become increasingly used in recent times. CNN models are used a lot in image recognition. Now, researchers are exploring best practices in applying this knowledge to financial data to find out the pattern of image data, and as such create predictive models from these patterns. From (Chen et al., 2021), convolutional neural networks were used on graphs to predict the future outlook based on the current present prices. The financial data was first converted into graphs then the graphs which are now our image data are placed into the CNN algorithm for learning and forecasting. The CNN model was also implemented on Forex data by (Hu et al., 2021) in addition to other models. The traditional candlestick data was used for the image data and simulated. Currently, the most common image data for CNN prediction is the candlestick graph. Moghaddam (Moghaddam & Momtazi, 2021) used this traditional candlestick method in his Forex analysis. There has been a strong argument that the CNN models do outperform other simpler models. However, the data characteristics also play a very important role in determining the best model. (Sinha et al., 2022)

This paper uses the S&P500 to model these three metrics and attain which of these models is the best for prediction. It further describes how the data characteristics interact with each of these models and then finds a fitting conclusion to each of the models.

## 2. METHODOLOGY

The simplest method of forecasting is the simple AR method. Its equation is derived from the interaction between the previous values and the present current value of a stock. This equation produced assists us in finding the future value at t+1, which in turn will be used to calculate the value of the stock at t+2 and the loop continues on and on.

$$\hat{Y}_t = A_1 Y_{t-1} + A_2 Y_{t-2} + A_3 Y_{t-3} + \cdots + A_p Y_{t-p} \qquad \text{1a}$$

$$\hat{Y}_t = \sum_i^p A_i Y_{t-i} \qquad \text{1b}$$

From Equation 1, we have arrived at conclusive evidence of the current price Y of the stock at time t which is a function of the prices from time t-1 to time t-p where p is a finite positive real integer. From modeling techniques, we know that a random error exists between the calculated and the original reading. (Kristjanpoller & Minutolo, 2015)

$$Y_t = \varepsilon_o + A_1 Y_{t-1} + A_2 Y_{t-2} + A_3 Y_{t-3} + \ldots \ldots + A_p Y_{t-p} \qquad \text{2a}$$

$$\boxed{\hat{Y}_t = \sum_i^p A_i Y_{t-i}} \qquad \text{2b}$$

Equation 1 and 2 shows a simple autoregressive model with order p, i.e. the current value depends on the p values in the past. The difference between the two equations is the error term which separates the predicted readings from the actual value.

Another metric used in trading is the moving average model. The raw model itself is usually used alongside AR models to derive maximum outputs.

$$\boxed{M.A = \frac{Sum\ of\ prices\ of\ q\ consecutive\ days}{q}} \qquad 3$$

Equation 4 represents the combination of the AR and the MA model which cumulates in the ARIMA model. The first part of the summation in equation 4 represents the AR and the second summation represents the MA model. In bridging the gap between the error terms, a moving average model was integrated into the AR model. From equation 4, $e_t$ is the independent and identically distributed variable with a 0 mean and standard deviation of 1. The MA model could model the random error to obtain an average pattern which would assist the model to predict even more accurate values.

$$Y_t = \mu + \sum_{i}^{p} A_i Y_{t-i} + \sum_{i}^{q} B_i \varepsilon_{t-i} \qquad 4$$

In considering the GARCH/ARCH model, we estimate the volatility which is of two types: conditional volatility and unconditional volatility. Conditional volatility or conditional variance is the variance of the returns of a stock given a measure. From measure theory, we can define conditional volatility as the variance given the σ-algebra. Unconditional volatility does not have such measure space constraints. In measurement, we need a standardized metric, so the log returns have been commonly used in forecasting. Take a stock Y with price P. The log return for the stock is

$$X_t = \log P_t - \log P_{t-1} = \log\left(\frac{P_t}{P_{t-1}}\right) \qquad 5$$

For the values of $X_t$, the unconditional volatility ($\sigma t$) can be calculated as

$$\sigma_t = \sqrt{Var[X_t^2]} \qquad 6$$

And the conditional volatility as

$$\sigma_t = \sqrt{Var[X_t^2 \mid F_t]} \qquad 7$$

Where $\mathcal{F}_t$ is the σ-algebra generated by the random variable X.

ARCH MODEL

$$\sigma_t = \sqrt{\omega + \alpha_1 X_{t-1}^2 + \ldots + \alpha_p X_{t-p}^2} \qquad 8$$

$$X_t = e_t \sigma_t \qquad 9$$

$$\therefore X_t = e_t \sqrt{\omega + \alpha_1 X_{t-1}^2 + \ldots + \alpha_p X_{t-p}^2} \qquad 10$$

Generalizing the ARCH model for conditional variance, we obtain GARCH MODEL.

$$\sigma_t \sqrt{\omega + \alpha_1 X_{t-1}^2 + \ldots + \alpha_p X_{t-p}^2 + \beta_1 \sigma_{t-1}^2 + \ldots + \beta_p \sigma_{t-p}^2} \qquad 11$$

Substituting from equation 9, we get:

$$\sigma_t = \sqrt{\omega + \alpha_1 X_{t-1}^2 + \ldots + \alpha_p X_{t-p}^2 + \beta_1 \sigma_{t-1}^2 + \ldots + \beta_p \sigma_{t-p}^2} \qquad 12$$

VAR MODEL

$$Y_t = K + A_1 Y_{t-1} + A_2 Y_{t-2} + A_3 Y_{t-3} + \ldots\ldots + A_p Y_{t-p} \qquad 13$$

In which A is the coefficient matrix of order p by t-p And Y is the matrix of order t-p by 1

## 2.1 THIEL'S U2 STATISTIC

Thiel's U2 statistic is the preferred measurement statistic for measuring the accuracy of a forecast. It compares the forecast to the benchmark (naïve) method. This statistic assumes the 1-step ahead method.

$$U_2 = \sqrt{\frac{\sum_{t=1}^{f} \left(\frac{\hat{y}_t - y_t}{y_{t-1}}\right)^2}{\sum_{t=1}^{f} \left(\frac{y_{t-1} - y_t}{y_{t-1}}\right)^2}} \qquad 14$$

$$\begin{cases} U_2 > 1 & \forall \text{ The forecast is worse than guessing} \\ U_2 = 1 & \forall \text{ The forecast is equal to guessing} \\ U_2 < 1 & \forall \text{ The forecast is better than guessing} \end{cases}$$

## 2.2 ASSUMPTIONS OF THE STUDY

A. **Data Completeness and Quality:** Foremost, all models require quality, complete, and clean data for robust results. This paper ensured that all the data input were free from all missing values, outliers and noise. Furthermore, since the model is being trained on time-series data, according to the paper, the data was standardized using the log-returns as the input and the stationarity was confirmed using the Augmented Dickey Fuller test (ADF-Test)

B. **Model Fit:** The second assumption is that each model has been optimized for its parameters and is fit for the specific data being analysed. For each of the models, different iterations were run based on varying hyper-parameters and the best fit to the data distribution which was viewed through the probability diagrams.

C. **Market Efficiency:** The analysis assumes semi-strong form efficiency, meaning current equity prices already reflect all publicly available information.

D. **Stability:** The structural relationships in the VAR model and the volatility patterns in the GARCH-ANN model remain stable throughout the forecast horizon.

E. **Features Representativeness:** For 3D-CNN models, it's assumed that the features extracted are representative of the underlying factors affecting equity prices.

F. **Nonlinearity and Volatility Clustering:** The GARCH-ANN model assumes that financial time series exhibit nonlinearity and volatility clustering.

G. **Model's Inherent Assumptions**: Intrinsic assumptions of each model, like the residuals being white noise in VAR, are maintained.

## 2.3. DATA SUMMARY

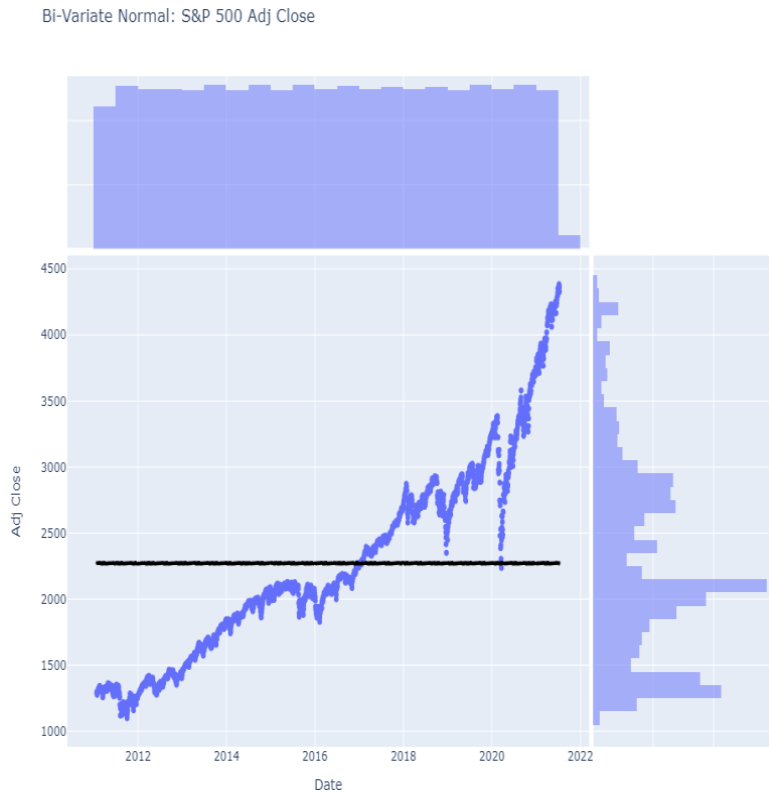

*Figure 1 Bi-Variate Normal: S&P500 ADJ CLOSE*

The data used was the S&P normalized collection of 500 traded US assets. The data recorded were the adjusted closing prices of the S&P500 dated from Friday, January 21, 2011, to Wednesday, April 20, 2022, taken daily, excluding weekends and public holidays, in which the equity markets are closed. The financial data was divided into training and testing sets, which were used to derive the various forecast equations for each of the test models under consideration and were used to test the accuracy of the models respectively. The total amount of data collected was 2,831 values. The data was further split into training and tests in the ratio of 2639 to 192 values respectively which is a 93% to 7% ratio. The entire work was simulated using the python programming language.

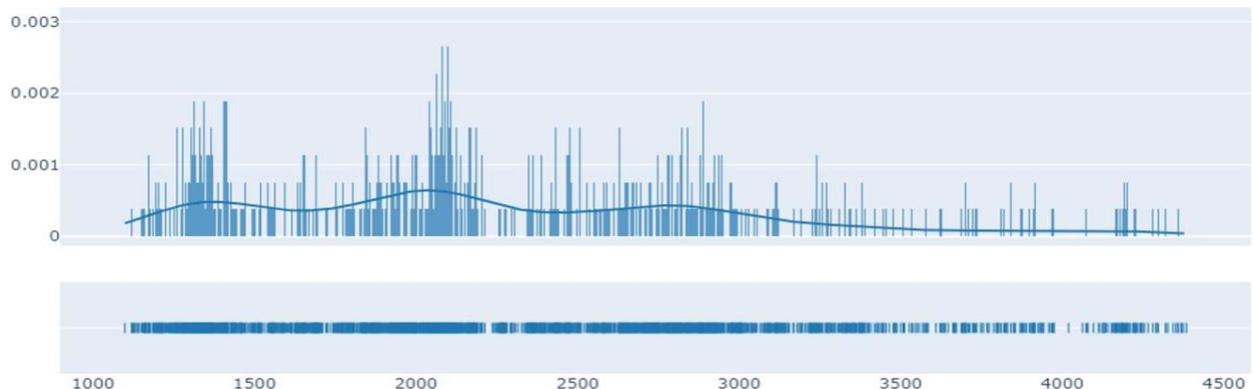

*Figure 2 Distribution of the S&P500 Price*

In figure 1, a bivariate normal curve that consists of the date and price distribution is shown to describe the data points of the training set. The top histogram shows us the date distribution every year. The reason for the discrepancies of one or two days is evident in the bars because, in each year, the number of market open days varies because of additional holidays, increased weekends, or even leap years. The right histogram shows the price distribution in USD of the S&P 500, with the mode prices ranging between $2,050 and $2,100 for 288 market days.

This distribution allows us to class the data, and therefore understand a mean reversion in price. The graph shows us the entire training set values in a scatter plot. This represents the 2639 values that were used in the training simulation.
In figure 2, a continuous probability distribution is used for the skewed data. The type of distribution is the gamma distribution. This distribution here shows the probability levels of each price class. Notably, the prices between $2000 and $2250 have the highest probability peaks.

After $3000, the curve fit seems to slope downwards. The reason why this graph might not give us the best intuition is that time-series data is non-stationary, thus, at every instant of time, we have a new value, except we assume that the price values are a result of only Brownian motion which is not fully true. Therefore, there is a need to analyze it in a non-stationary environment.

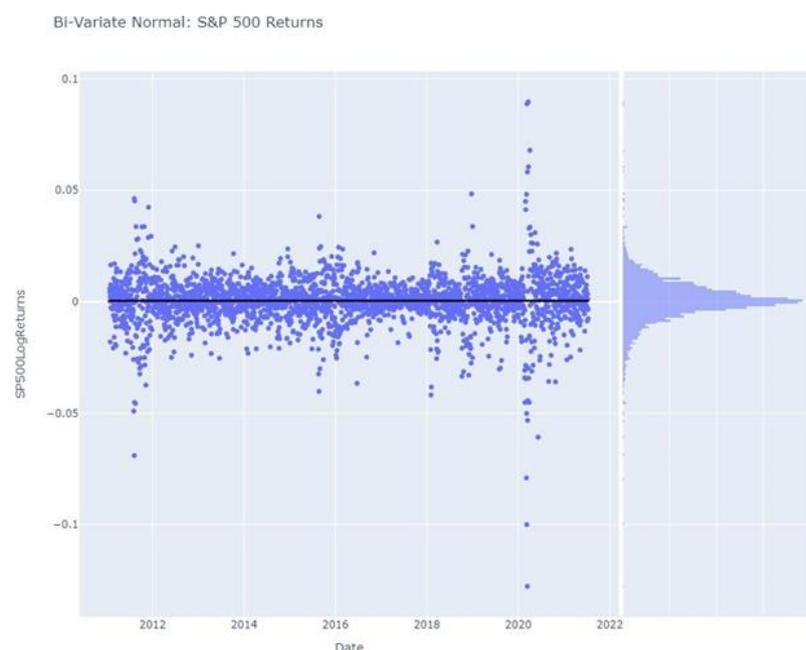

*Figure 3 Bi-Variate Normal: S&P 500 Returns*

In figure 3, the scatter plot of the returns is evident, and this produces a non-stationary effect. The date-data distribution was ignored as it is the same as the price distribution in figure 1, however, the log-returns show us a normalized graph. Therefore, the price distribution is a log-normal graph. From the distribution, the graph is negatively skewed with a skewness coefficient of -0.93 which means that it is more aligned to the positive returns.

Using a scale of 1:10000, the gamma distribution was scaled upwards to give a clear depiction of the returns, as the probability spread is evident in the normal distribution.

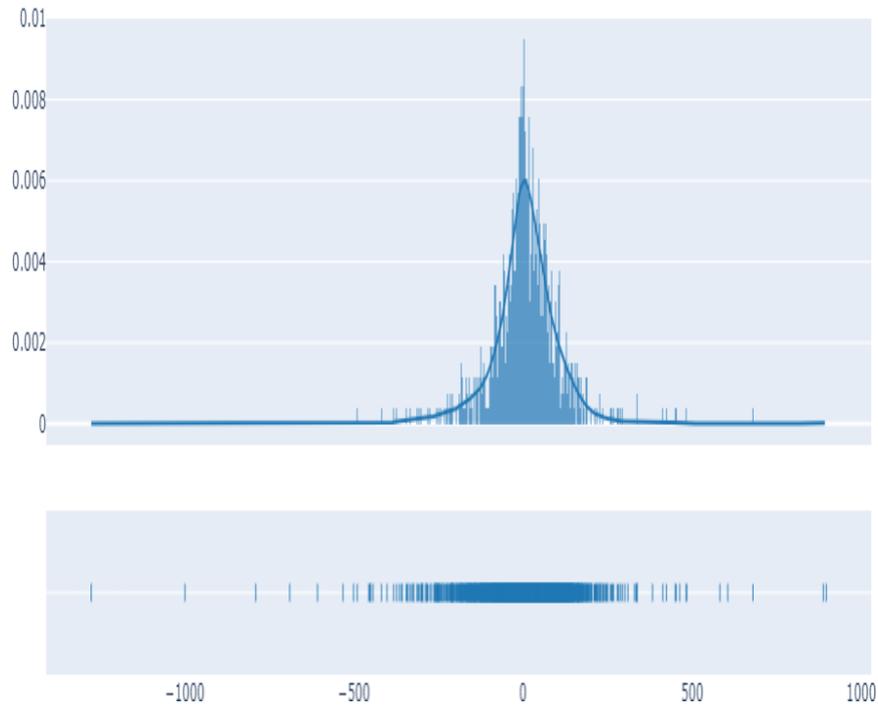

*Figure 4 Distribution of the S&P500 Returns*

TABLE 1: METRIC RESULTS OF THE S&P500 RETURN

| METRIC | VALUE |
|---|---|
| Mean | 0.000461 |
| Median | 0.000669 |
| Standard deviation | 0.010935 |
| Variance | 0.00012 |
| Minimum | -0.127652 |
| Maximum | 0.089683 |
| Kurtosis | 17.76447 |
| Skewness | -0.927768 |

# 3. MODEL STRUCTURE AND ARCHITECTURE

## 3.1 3D-CNN STRUCTURE

3D-CNNs, as the name implies, operate on three-dimensional data, which includes depth in addition to height and width. In the realm of financial data analysis, particularly for the S&P500, one might typically conceptualize the "depth" as various data indicators, such as closing prices, volume, or moving averages. However, in our unique approach, we're utilizing the visual representation of returns as the basis for the 3D structure.

By plotting the returns graphically, two dimensions are readily apparent: the height and width of the charted returns over time. The third dimension is introduced by the actual value of the returns, which adds depth. Together, these three dimensions—height, width, and depth of return values—form the basis of our 3D-CNN model. This creative approach leverages the inherent sequential and temporal patterns in return values to optimize predictive capabilities.

## 3.2 3-D CNN ARCHITECTURE:

**Input Layer:** The input layer processes sequences from our training dataset, where each day is represented by a tensor with dimensions corresponding to height, width, and depth. These dimensions encapsulate various financial indicators pertinent to our analysis.

**Convolutional Layers:** The model incorporates several 3D convolutional layers. Each convolutional layer utilizes a kernel of size 3x3x3 to scan the input tensor. The "relu" activation function is applied after each convolutional operation to introduce non-linearity into the model.

**Pooling Layers:** To reduce the spatial dimensions of our data while retaining critical features, 3D max-pooling layers are interspersed between convolutional layers. Each pooling layer operates with a pool size of 2x2x2.

**Dense Layers**: After several convolutional and pooling operations, the model employs dense layers to assist in the final stages of prediction. These layers help in learning intricate patterns and relationships from the spatially reduced features.

**Dropout Layers:** To mitigate the risk of overfitting given the complexity of our model, dropout layers are strategically placed within the network. They randomly set a fraction of input units to 0 at each update during training time.

**Output Layer:** The final dense layer outputs a singular value representing the predicted closing price for the next day.

## 3.3 TRAINING DETAILS OF THE 3-D CNN:

**Dataset:** Our model was trained on S&P500 data spanning from Friday, January 21, 2011, to Thursday, July 15, 2021. This timeframe provided 2,639 data points for training and validation. This sa

**Training Epochs:** We subjected our model to 100 training epochs, ensuring an adequate balance between convergence and computational efficiency.

**Framework and Libraries:** The entire training and model architecture were implemented using the TensorFlow package, leveraging its high-level API, Keras, for streamlined model development.

## 3.4 GARCH-ANN/LSTM STRUCTURE

The model weaves together the strengths of the GARCH (Generalized Autoregressive Conditional Heteroskedasticity) approach, renowned for its prowess in volatility forecasting, with the capabilities of an Artificial Neural Network (ANN) tailored for price prediction. Predictions are made for the returns using both techniques, after which respective confidence weights are assigned to each forecast. These weighted forecasts are then synergistically combined for an integrated prediction.

## 3.2 GARCH-LSTM ARCHITECTURE:

**GARCH Component:** The primary function of the GARCH component is to capture the time-varying volatility in financial time series data. Volatility clustering, where periods of high volatility are followed by similar periods and likewise for low volatility, is a common phenomenon in financial markets.

By modeling the conditional variance, GARCH predicts future volatility based on past return data. These volatility predictions can be crucial, especially when the focus is on risk management.

**LSTM Component:** The ANN is primarily responsible for predicting returns or prices based on the past values and other potential features. Neural networks, with their ability to capture non-linear relationships and intricate patterns, provide a sophisticated tool for this purpose.

Typically, the network would be fed with past return values (and potentially other features) and trained to predict the next period's return or price.

**Input Layer:** The input layer for this model takes sequences of a rolling 30 days' closing prices. Then it will predict the price on the 31st value. In other words, it uses 30 values as input to determine the next output. Each day's closing price is a single value, forming sequences with a shape of (30, 1). This layer essentially sets the stage for the data to flow into subsequent layers in the network. This continues through the entire training dataset until it forms an approximate formula that can capture the dependency of the data

**LSTM Layers:** The LSTM layers in tensorflow act analogous to convolutional layers in CNNs. The first LSTM layer consists of 50 units. Each unit aims to capture temporal dependencies within the input sequences. It is set to return_sequences=True, meaning it will return the full sequence to the next layer, allowing for the stacking of multiple RNN layers.

The second LSTM layer, also with 50 units, processes the sequences from the prior layer and condenses the information, getting the data ready for the subsequent dense layer. The choice of 50 was based on the trade-off of complexity and computational space.

**Dense Layers:** The model has one dense layer with a single unit. This layer is responsible for generating the final output based on the features extracted by the LSTM layers. The unit produces a singular value, predicting the next day's closing price.

**Output Layer:** Technically, in this code setup, the single unit dense layer also serves as the output layer. The model employs the Adam optimization algorithm to minimize the mean squared error, which measures the difference between the model's predictions and the actual values. Given the sequential and temporal nature of the stock price data, the use of LSTM layers makes intuitive sense as they are designed to capture patterns over time.

**Integration:** After obtaining the forecasts from both the GARCH and ANN components, the next challenge lies in integrating these results. This is achieved by:

**Weighting Mechanism:** Each forecast is assigned a confidence weight. This weight can be derived from historical accuracy, the model's intrinsic confidence measure, or other heuristic methods.

**Combination:** The weighted forecasts are then combined to produce a final prediction. This combination could be a simple weighted average or a more complex function depending on the design choices.

**Evaluation:** Consider both the accuracy of the price prediction (using metrics like RMSE) and the accuracy of the volatility forecast.

### 3.4 VAR STRUCTURE
VAR is a multivariate forecasting method that uses a system of regression equations, where each equation predicts a single time series as a function of past values of all-time series in the system.

### 3.5 VAR MODEL SPECIFICATION
**Endogenous Variables**: Both SP500 and SPDR1500 closing prices are considered as endogenous, meaning they depend on their own past values and the past values of each other.

**Lags**: The VAR model considers several past values (lags) of each time series. Using ACF and PACF lags, the 8th lag was the most suitable option for the log-returns. The number of lags is determined automatically based on information criteria (like AIC) during the model fitting phase.

**Equations**: Each variable has its own equation. For instance, for the GSPC, the forecasted value might be a linear combination of its past values, the past values of SPDR1500, and an error term. Similarly, there will be an equation for SPDR1500.

**Error Term:** The error term is vital, as it captures all other influences on the variable not covered by the model. It is assumed to have a mean of zero and is not correlated with past values of itself or other variables.

**Forecasting**: After the model is trained on the historical data, it's used to generate forecasts for the duration of the test set. These forecasts are based on the most recent data points (up to the specified lag order) from the training dataset.

**Stationarity Check:** Before estimating a VAR model, an Augmented Dickey Fuller test was performed to determine the stationarity of the data and it was confirmed that the data is stationary. This step is important because VAR models do not do well with non-stationary data. This involves checking its statistical properties like its mean and variance. The stationarity check was also carried out on SPDR1500 for the same time period of forecasting.

## 4. RESULTS

To compare properly the forecast accuracy, we use different error metrics namely:

- Mean Absolute Percentage Error (MAPE)
- Mean Error (ME)
- Mean Absolute Error (MAE)
- Mean Percentage Error (MPE)
- Root Mean Square Error (RMSE)

### 4.1. GARCH – LSTM

The LSTM modeled the conditional volatility of the GARCH model. This combination brought out a better performance compared to the original GARCH model. From the structure of the returns which was viewed using the partial autocorrelation and the autocorrelation correlogram, the p, q coefficient was recorded as 1, 4, with a normal distribution. From studies on Brownian motion by L. Bachelier, financial data does not depend on previous values, however, we have noticed over studies that this is not entirely the case, thus the need for the long short term memory (LSTM) which was used to model the conditional variance.

From figure 5, the GARCH model does indeed produces a strong replication of the original graph. It modeled the positive peak periods well but seems a little biased on the negative peaks.

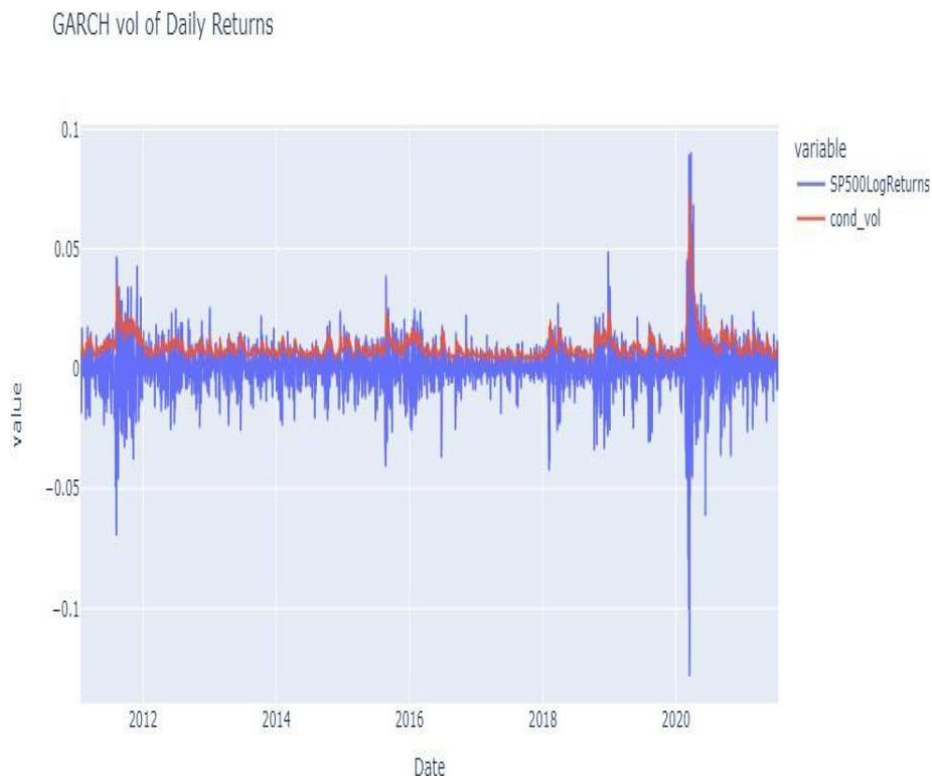

*Figure 5 GARCH Simulation of the Train Dataset*

### 4.2. VAR Model

The VAR modeled the log returns of the S&P500 using the relationship between the log returns of the SPDR1500 (Standard & Poor's Depositary Receipt 1500) and its returns which exhibited a high correlation. The price correlation was approximated at 99.962%, while the returns correlation was approximated at 96.518%. Since the assumption was just to use two highly correlated stocks, the SPDR1500 did a good job of that. The next step was finding the best fit value for the order of the VAR model. Usually, those with the lowest AIC fit the bail, as the VAR model is repeated with increasing orders.

*Table 2 VAR Regression Analysis*

| Summary of Regression Results | |
|---|---|
| Model: | VAR |
| Method: | OLS |
| No. of Equations: | 2 |
| BIC: | -21.1288 |
| Nobs: | 2629 |
| QIC: | -21.183 |
| Log-likelihood: | 20462.7 |
| FPE: | 6.12E-10 |
| AIC: | -21.2137 |
| Det(Omega_mle): | 6.04E-10 |

To choose the proper order of the VAR model, we repeatedly fit increasing orders of the VAR model and choose the order that produces the model with the lowest AIC (Akaike information criterion). Table 2 shows the fitting criterion of the dataset.

### 4.3. 3-D CNN MODEL

Like the other models, the log-returns were simulated. The idea was to create a simple convoluted matrix that is representative of the different images that appear on a typical graph. The matrix was divided into image rows and image columns and a fake depth channel were added to provide omitting data points folded at the edge.

### 4.4. RETURNS COMPARISON

After using the three models, each of them brought forth their simulated return readings with 3D CNN and the VAR model trying to converge at every time. The error measurement of the metrics will provide a solid ground for comparison.

The scatter plots of the simulated returns are given in figure 6. The correlation of the GARCH-LSTM was rounded at 1.88%, with the positive part of the points more closely simulated than the negative points, however, the outliers were not properly simulated in the forecast model.

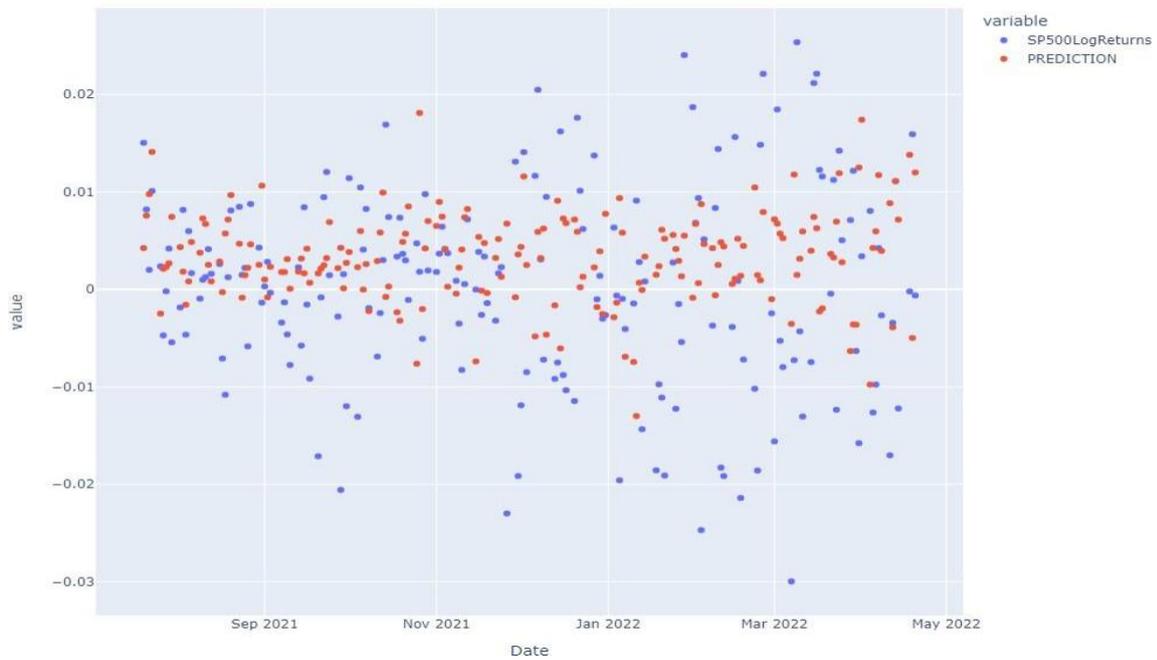

*Figure 6 GARCH-LSTM SCATTER PREDICTIONS OF STOCK RETURNS*

The VAR model itself recorded a negative 6.49% correlation which could spell errors for the return series. The 3D-CNN model scatter-plot of returns is shown in figure 7, where the readings follow a uniform trend. Notably, the log prices of the plot are well spread and still follow the normal distribution, however, it fails to properly model any value above and below ±1% respectively.

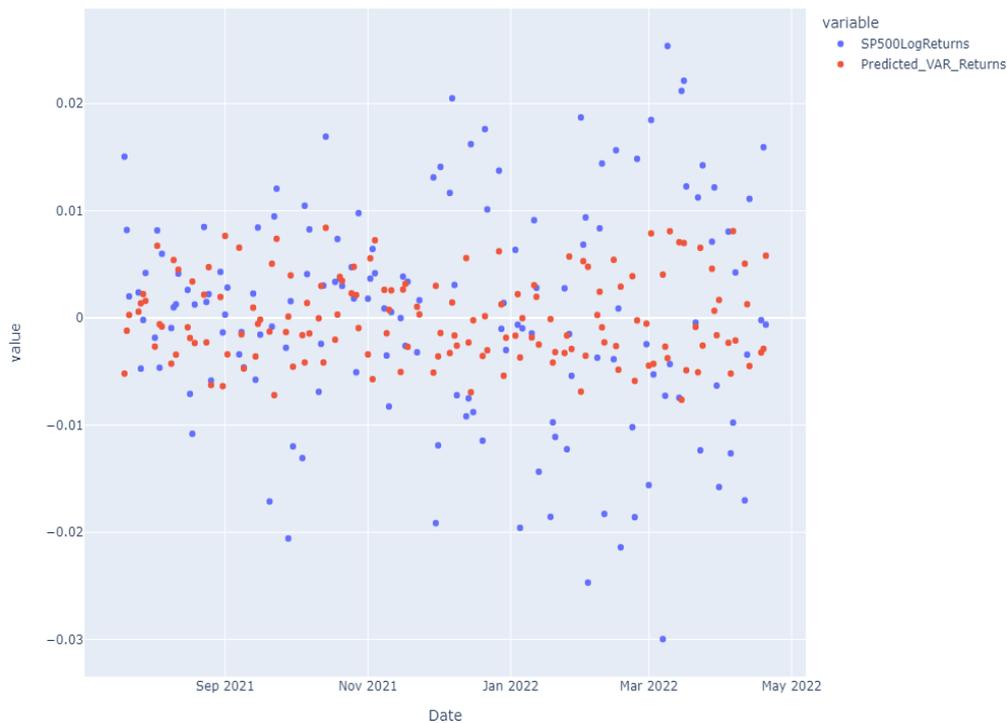

*Figure 7 VAR SCATTER PREDICTIONS OF STOCK RETURNS*

When viewing the returns alone, we can assume that the VAR produces the least error metric because it has the least error metric. However, for better comparison, the prices will be a good place to start.

**4.5 PRICE PREDICTIONS**

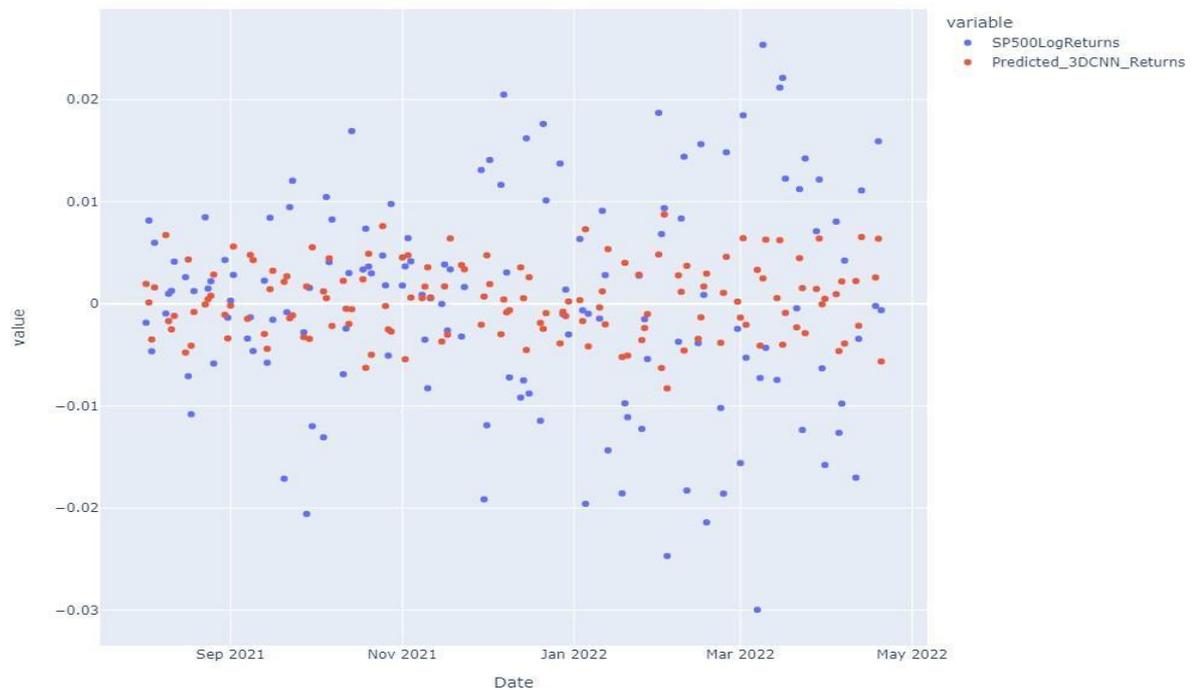

*Figure 8 3D-CNN SCATTER PREDICTIONS OF STOCK RETURNS*

The forecast overall aims to provide a strong pricing model which will assist the decision-making of the average investor. The GARCH price prediction data is represented in the graph in figure 9. The volatility captured in the GARCH model appears to be well-positioned as is natural with financial data. The model begins fairly well, but the spread increases with increase in the months, especially within October 2021, and then February to March 2022, where we see a downward trend whereas our model did predict a downward trend, the original trend was far steeper and as a result, caused the model to be weakened.

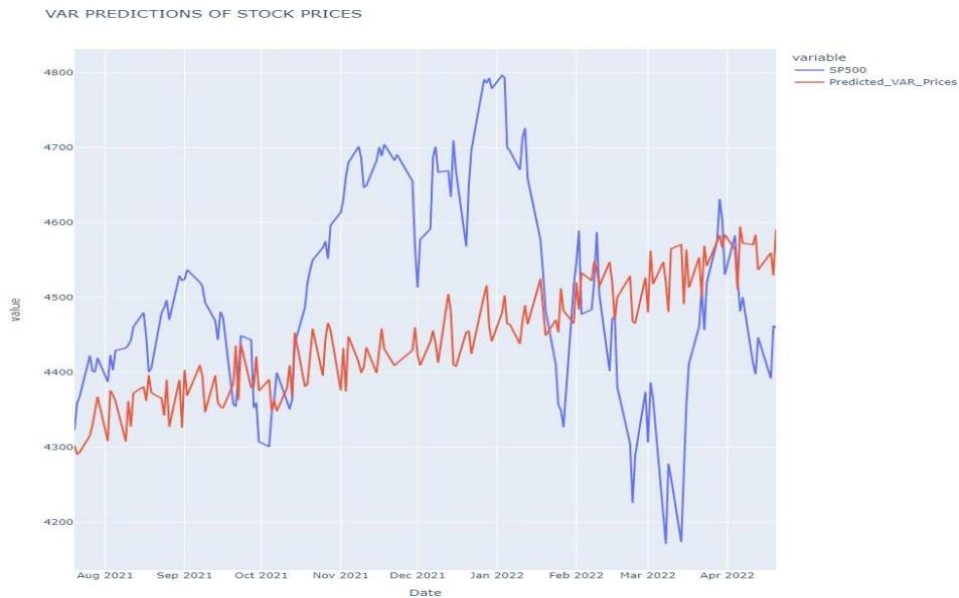

*Figure 9 GARCH-LSTM Predicted Price Movement*

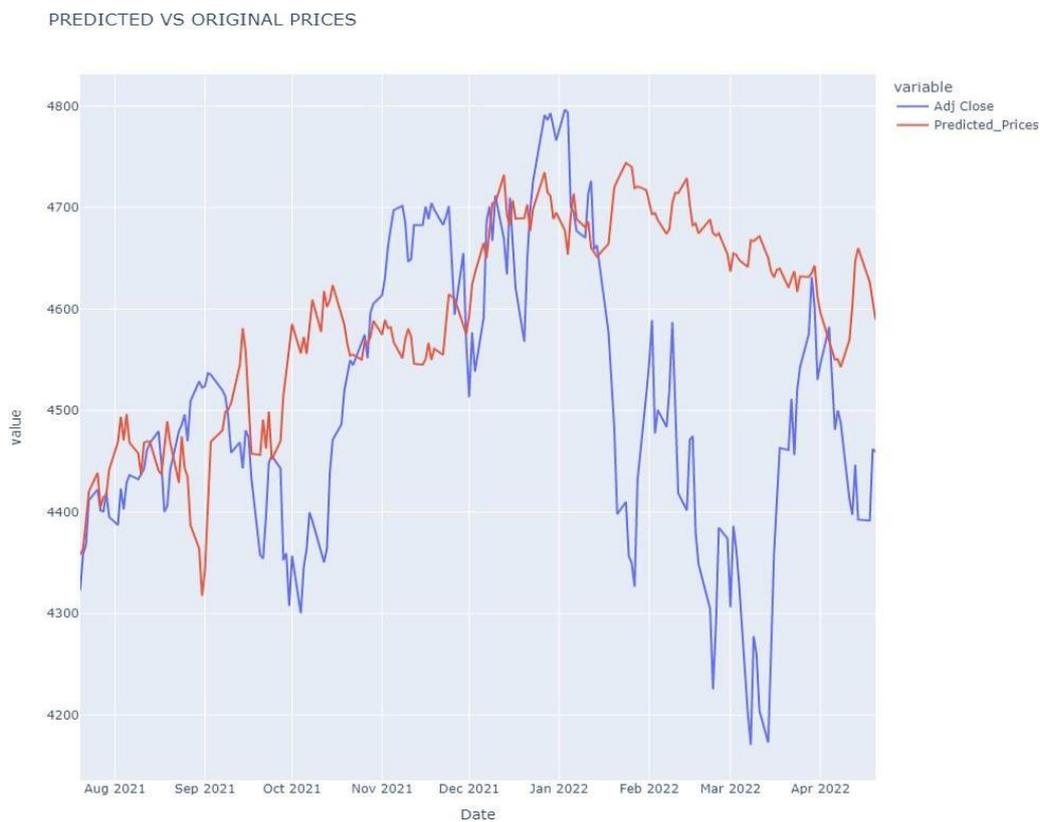

*Figure 10 VAR MODEL PREDICTED PRICES*

Figure 10, shows the VAR model simulation of the predicted prices. The volatility is quite extreme as the algorithm worked with simulating the random error to give a stronger prediction. The trend was somewhat like an upward followed by a downward trend, which makes the returns quite oscillatory. It is notable that during the period where the GARCH-LSTM failed, the VAR molly gave a closer accurate prediction, however, the fall in March was not anticipated by the model

*Table 3 LOG PRICES ERROR METRICS*

| METRIC | GARCH-LSTM | VAR | 3D-CNN | BEST |
|---|---|---|---|---|
| MEAN ABSOLUTE PERCENTAGE ERROR (MAPE) | 2.3366 | 4.8144 | 3.7593 | GARCH-LSTM |
| MEAN ERROR (ME) | 2.26E-05 | -3.56E-04 | 1.92E-04 | GARCH-LSTM |
| MEAN ABSOLUTE ERROR (MAE) | 0.009308912 | 0.008649 | 0.008503 | 3D-CNN |
| MEAN PERCENTAGE ERROR (MPE) | -0.805683556 | 2.266984 | 1.398464 | GARCH-LSTM |
| ROOT MEAN SQUARE ERROR (RMSE) | 0.011490963 | 0.011215 | 0.010919 | 3D-CNN |

*Table 4 ERROR METRICS FOR FORECASTED PRICES OF THE S&P 500*

| METRIC | GARCH-LSTM | VAR | 3D-CNN | BEST |
|---|---|---|---|---|
| MEAN ABSOLUTE PERCENTAGE ERROR (MAPE) | 0.029087551 | 0.030228 | 0.033357 | GARCH-LSTM |
| MEAN ERROR (ME) | 85.80646155 | -56.51107 | 32.5675 | 3D-CNN |
| MEAN ABSOLUTE ERROR (MAE) | 128.6399751 | 136.9949 | 149.176 | GARCH-LSTM |
| MEAN PERCENTAGE ERROR (MPE) | 0.019879466 | -0.011621 | 0.008246 | 3D-CNN |
| ROOT MEAN SQUARE ERROR (RMSE) | 174.2774947 | 176.1822 | 182.0202 | GARCH-LSTM |
| CORRELATION | 0.218999302 | 0.040342 | -0.05237 | GARCH-LSTM |
| MINMAX | 0.027703295 | 0.029824 | 0.032156 | GARCH-LSTM |

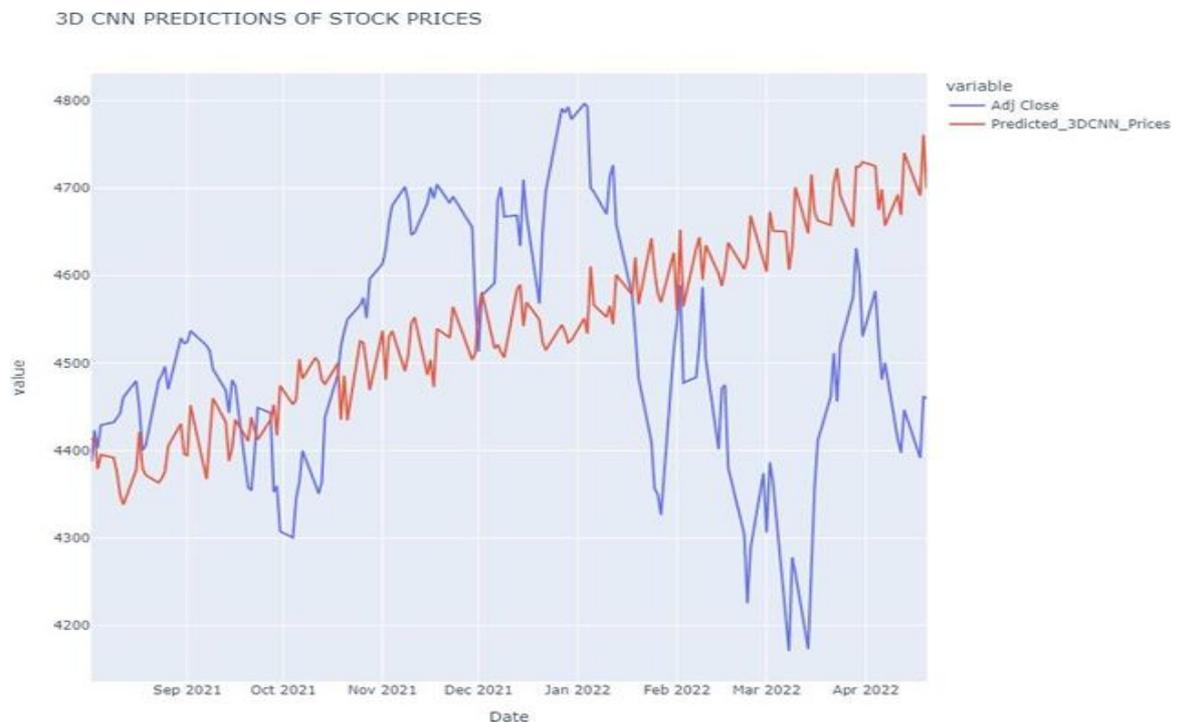

*Figure 11 3D CNN PREDICTION OF THE STOCK PRICES* From the graph of the prediction of the 3D-CNN, we can assume that the stock prices can eventually rise infinitely. This prediction came as a result of the returns trying to converge to produce a unified pattern, which is not possible in reality. This means that the simple 3D-CNN does not perfectly provide for the complexities of everyday volatility which has caused the model to appear like this.

Overall, the GARCH-LSTM model was the one with the highest correlation figure at 22% and the least root mean square error among the three models.

The GARCH-LSTM model can be considered an "academic cross-pollination," synthesizing elements of time-series econometrics with recurrent neural networks. The GARCH component captures volatility clustering—a phenomenon prevalent in financial data—while the LSTM layers provide the model the ability to learn from long-term sequential dependencies. This duality may account for its superior RMSE and MAPE scores, making the model both an accurate and unbiased predictor.

Interestingly, the model does not dominate across all metrics. Specifically, the 3D-CNN model presents a slightly better MAE score, indicating that the complexity of GARCH-LSTM may sometimes capture noise as signal. This raises an important question: does the drive for higher complexity necessarily improve real-world forecasting, or does it, at some point, start to introduce 'academic overfitting'?

Another avenue of interest is the intra-model synergy between the GARCH and LSTM components. How do these different methodologies interact at a granular level to produce a model that outperforms its components in isolation? We propose that the LSTM component may be acting as a 'dynamic recalibrator' for the GARCH model, adapting its predictions in real-time to account for unforeseen structural breaks or anomalies.

## 5. CONCLUSION AND FUTURE WORK

The importance of forecasting in finance cannot be overemphasized. The models together did a good job of forecasting the increase. In 192 days, the S&P growth should have been 4.72%, however, the GARCH-LSTM model forecasted a 5.17% growth with a U2 statistic of 1.7. This shows that the model would not necessarily have done better than guessing, however, seeing that the period is a long stretch and the GARCH-LSTM model could model varying levels of volatility, it could be a guiding point for investors. The GARCH-LSTM performed the best amongst the 3 models, with the least mean absolute percentage error, the least mean absolute error and the least root mean square error. The 3D-CNN which underperformed among the three models recorded a growth forecast of 7.72%. This could result in a serious error when offering investment advice. The U2 statistic of the 3D-CNN was 3.9, which shows that the path of the model did not perform as well as we would have loved it. The model has the least error value when evaluating the mean error and the mean percentage error. Here guessing is more suitable. The third model in our study which had an average spread of error recorded a U2 statistic of 3.5 and a growth rate spread of 5.4%. This model came in second and had the second-highest root mean square error which also tallied with the positions of Thiel's statistic.

In our study, we found out that it was fairly obvious why the GARCH-LSTM model will perform better, because it is a complex model, while the VAR and the 3D-CNN were simple models. However, the 3D-CNN model has become increasingly popular in recent times. To simulate stronger readings from this model, it has been noted that the model should have many other metrics which are usually available in candlesticks like the minimum price, maximum price, and the adjusted close for each day which will help the algorithm simulate a better reading.

Overall, our GARCH-LSTM model outperformed the other two models for forecasting mid-term. In future works, the GARCH-LSTM model will be improved with greater learning time to improve the accuracy, also including other financial metrics like the daily low price, daily high price, the close prices, the adjusted closing prices, daily volatility metric, and the market open prices. The test will be to see if indeed these metrics can further improve the accuracy of the GARCH-LSTM and even further improve the accuracy of the other three models and if indeed, they will exhibit the same behavior.

Although the GARCH-ANN and VAR models were trained on the same dataset, their parameters can't be regarded as constants every time they're trained. The unique nature of the GARCH-ANN model, being a hybrid that combines both statistical and neural network elements, ensures that its hyperparameters will differ from a purely statistical model like VAR. Furthermore, the GARCH component of the hybrid model factors in volatility in its calculations, an element not directly considered by the VAR model. This difference in model structure and parameter consideration implies that the parameters for these models won't always be consistent, even when trained on identical data.

# 6. LIMITATIONS OF THE STUDY

**Historical Dependency:**
All models rely on historical data, which might not always be indicative of future prices, especially during unprecedented market shocks.

**Model Complexity:**
3D-CNNs can be computationally intensive, requiring more computational power and time to train than GARCH-ANN or VAR.

**Overfitting:**
There's a potential risk that models, especially deep learning ones like 3D-CNN, might overfit the training data, leading to poor generalization on new data.

**Model Specificity:**
Each model has its inherent limitations. For instance, VAR models are linear and might not capture complex nonlinear relationships. Similarly, GARCH-ANN assumes certain patterns of volatility which might not hold in all scenarios.

**External Factors:**
There are external and exogenous factors, such as policy changes, geopolitical events, etc., that can influence equity prices and may not be captured entirely by the models.

**Lack of Interpretability:**
3D-CNN models can act as black boxes, making it difficult to infer which features are the most influential in price prediction.

**Horizon Limitation:**
Predicting equity prices further into the future can be challenging and less accurate due to the increased uncertainty.

Conflict of Interest

There is no conflict of interest whatsoever.

Author Contribution

Conception: [SA, MIA, CYB]

Design: [SA, MIA]

Execution: [SA, MIA]

Interpretation: [SA, MIA, CYB]

Writing the paper: [SA, CYB]